# Defect configuration, not nitrogen content, governs the mechanical integrity of nitrogen-doped graphene: a molecular dynamics study

Indranil Rudra [1], Jahid Emon *[2], A.K.M. Monjur Morshed [3]

[1] Department of Mechanical and Aerospace Engineering, The University of Alabama in Huntsville, AL 35899, United States

[2] Department of Mechanical Science & Engineering, University of Illinois Urbana-Champaign, Urbana, IL 61801, United States

[3] Department of Mechanical Engineering, Bangladesh University of Engineering and Technology, Dhaka, Bangladesh

* Corresponding author: jemon2@illinois.edu (Tel: 447-902-7621)

## Abstract

The mechanical reliability of nitrogen-doped graphene is often attributed to its nitrogen content, yet nitrogen occurs in chemically distinct configurations whose individual mechanical roles, and whose interactions with other defects, remain unresolved. Here, molecular dynamics simulations of uniaxial tension are used to separate the contributions of nitrogen chemistry, missing atoms, and defect arrangement to the strength and fracture of graphene. Three size-matched defects, a graphitic-nitrogen cluster, a void, and a pyridinic-nitrogen cluster, are compared so that two controlled contrasts isolate the effects of edge chemistry and of the vacancy independently. The graphitic cluster leaves the mechanical properties essentially unchanged (a strength reduction of <1 %), whereas the void degrades the ultimate strength by ~23 % and the pyridinic cluster, which combines the same vacancy with edge nitrogen, is the most damaging (~30 %), failing abruptly from its nitrogen-decorated rim rather than through the damage-tolerant process of the bare void. The mechanical impact of a nitrogen cluster is therefore governed by whether it carries vacancies, not by nitrogen itself. When a nitrogen cluster and a void coexist, their interaction is controlled by orientation relative to the load: in-line defects interact negligibly and fail at the more severe member, whereas side-by-side defects couple through overlapping stress fields and weaken the sheet progressively as they approach, an interaction that persists to separations of ~80 Å. These results establish that the mechanical integrity of nitrogen-modified graphene is determined by the configuration of defects, the bonding environment of nitrogen, and the arrangement of coexisting defects relative to the load, rather than by nitrogen content or defect density alone, thereby providing a basis for defect-tolerant design.

## 1. Introduction

Graphene, a single atomic layer of $sp^2$-bonded carbon, exhibits a combination of mechanical properties unmatched by any other material. Direct nanoindentation measurements established its intrinsic Young's modulus at approximately 1 TPa and its breaking strength at approximately 130 GPa, identifying it as the strongest material ever measured [1]. These properties underpin a broad range of prospective applications, from ultra-strong structural composites to durable, flexible components for electronics and energy storage, in which graphene must bear or transmit mechanical load reliably [2].

In practice, however, the mechanical performance of graphene is dictated less by the ideal lattice than by the structural defects it inevitably contains. Defects are introduced unavoidably during growth, transfer, and processing, and they can also be deliberately engineered; in either case, they strongly modify the mechanical response, and their controlled introduction can even tailor graphene's fracture behaviour for specific applications [3]. A substantial body of atomistic work has consequently characterized how individual defect types govern strength and fracture. A single vacancy alone reduces the fracture strength of graphene by approximately 18 %, with the crack nucleating through the destruction of the carbon rings adjacent to the vacancy [4]. Grain boundaries weaken the sheet in a manner that depends on their structure: the intrinsic strength of flat grain boundaries decreases linearly with formation energy up to a maximum of about 93 GPa, while inflected boundaries weaken further with increasing inflection angle [5]. Holes and nanopores act as stress concentrators whose severity depends on their size and edge morphology, and a recurring theme across these studies is that the mechanical penalty of a defect is governed by how it perturbs the local stress field, by defect shape and edge structure, not defect density alone [6].

Nitrogen is among the most important substitutional dopants in graphene, widely used to tailor its electronic and chemical behaviour, and its mechanical consequences have accordingly received attention. Molecular dynamics simulations have reported that nitrogen doping can reduce the tensile strength of graphene substantially, by more than 35 % at a nitrogen concentration of only 2 %, with point vacancies further lowering both strength and stiffness [7]. In real samples, however, nitrogen is incorporated in several distinct bonding configurations, graphitic, pyridinic, and pyrrolic, whose proportions depend on synthesis. These configurations differ fundamentally in local atomic structure: graphitic nitrogen substitutes for carbon within the intact lattice, whereas pyridinic and pyrrolic nitrogen are associated with vacancies and under-coordinated edges. Because studies such as the above couple nitrogen substitution with vacancy formation, the extent to which these chemically distinct configurations differ purely mechanically, and whether the mechanical effect of a nitrogen cluster originates from the nitrogen chemistry itself or from the vacancies that accompany it, has not been systematically isolated.

A further and largely open question concerns defects that coexist. Real defective graphene contains many defects in proximity, whose perturbed stress fields may overlap and interact, yet the collective mechanical behavior of interacting defects has been studied far less than that of isolated ones. Existing work emphasizes that the stress field, rather than defect count alone, governs the collective response, and that grouping defects can homogenize the local stress and under some conditions even raise the strength relative to isolated defects [8]. This indicates that the mechanical interaction between defects must depend not only on their separation but on their geometric arrangement, including their orientation relative to the applied load, a dependence that has not been systematically quantified for chemically distinct coexisting defects.

The present work addresses these two gaps using molecular dynamics simulations of uniaxial tension in defective graphene. First, to disentangle the mechanical roles of nitrogen chemistry and missing atoms, three size-matched defects, a graphitic-nitrogen cluster (chemistry without missing atoms), a void (missing atoms without chemistry), and a pyridinic-nitrogen cluster (missing atoms with edge chemistry), are compared directly, enabling two controlled contrasts that isolate the effect of the edge nitrogen and of the vacancy independently. Second, the mechanical interaction between a nitrogen cluster and a void is quantified as a function of both their separation and their orientation relative to the loading axis, establishing how coexisting defects couple and over what range. Together, these analyses show that the mechanical integrity of nitrogen-modified graphene is governed by defect configuration, whether nitrogen carries vacancies, and how coexisting defects are arranged with respect to the load, rather than by nitrogen content or defect density alone.

The remainder of this paper is organized as follows. Section 2 describes the atomistic model, the defect construction, and the simulation and analysis protocols. Section 3 presents the mechanical response of the isolated defects and their fracture mechanisms, followed by the interaction between coexisting defects. Section 4 discusses the implications and broader context, and Section 5 summarizes the principal findings.

## 2. Methodology

### *2.1 Atomistic model and simulation cell*

All molecular dynamics (MD) simulations were performed using the Large-scale Atomic/Molecular Massively Parallel Simulator (LAMMPS, 29 Aug 2024 release) [9]. A single-layer, defect-free graphene sheet was first constructed as an orthorhombic supercell with the zigzag and armchair crystallographic directions aligned with the x and y axes, respectively, using a carbon–carbon bond length of 1.42 Å. The reference sheet contained 11040 carbon atoms with in-plane dimensions of approximately 169.7 Å × 170.4 Å. Periodic boundary conditions were applied along the in-plane x and y directions, while the out-of-plane z direction was non-periodic with a vacuum spacing of ~50 Å to eliminate interactions between periodic images of the sheet. The lateral dimensions were chosen so that each defect was separated from its nearest periodic image by at least ~50 Å, ensuring the perturbed stress field around a defect did not overlap with that of its image. For the largest perpendicular defect separation, the cell was enlarged along the zigzag (x) direction to ~219 Å (14240 atoms) to preserve this clearance.

### *2.2 Defect construction*

Three structurally distinct, size-matched defects were introduced at the centre of the sheet, each occupying a circular footprint of approximately 7 Å radius. A circular void was created by removing the 60 carbon atoms closest to the defect centre, producing a bare-edged nanopore. A graphitic-nitrogen cluster was created by substituting the 30 carbon atoms closest to the centre that belong to a single graphene sublattice with nitrogen. Because the graphene lattice is bipartite, every atom's nearest neighbors lie on the opposite sublattice. Restricting the substitution to one sublattice ensures that no two nitrogen atoms are nearest neighbors, so that every nitrogen remains three-fold coordinated and bonded exclusively to carbon, yielding a spatially aggregated cluster free of N–N bonds.

A pyridinic-nitrogen cluster was created by first removing the same 60-atom hole as in the void and then substituting the under-coordinated carbon atoms on the pore rim with nitrogen. The resulting edge nitrogen atoms (21 in total) are predominantly two-fold coordinated, reproducing the pyridinic configuration, with a small number of adjacent (pyridazinic) sites occurring naturally at the rim. The electronic and magnetic consequences of these peripheral nitrogen topologies (pyridinic, pyridazinic, pyrrolic, and pyrazolic) have been examined separately using first-principles calculations [10]. Here, we focus on their mechanical consequences. This defect, therefore, combines a missing-atom (vacancy) character with edge-nitrogen chemistry.

This construction enables two controlled comparisons: the void and the pyridinic cluster share an identical hole and differ only in edge chemistry (carbon versus nitrogen), isolating the effect of the edge nitrogen, whereas the graphitic and pyridinic clusters differ in the presence of the vacancy, isolating the effect of the missing atoms. To probe the mechanical interaction between coexisting defects, configurations containing one nitrogen cluster and one void were generated with the two defects separated by an edge-to-edge gap of 5, 10, 15, 25 and 40 Å (extended to 60 and 80 Å for the perpendicular orientation to confirm convergence to the non-interacting limit). Two relative orientations were considered: parallel, in which the defects were offset along the loading (armchair, y) axis (in-line), and perpendicular, in which they were offset along the transverse (zigzag, x) axis (side-by-side). Initial configurations were generated with an in-house Python script and the resulting geometries were cross-checked with the atomsk package [11].

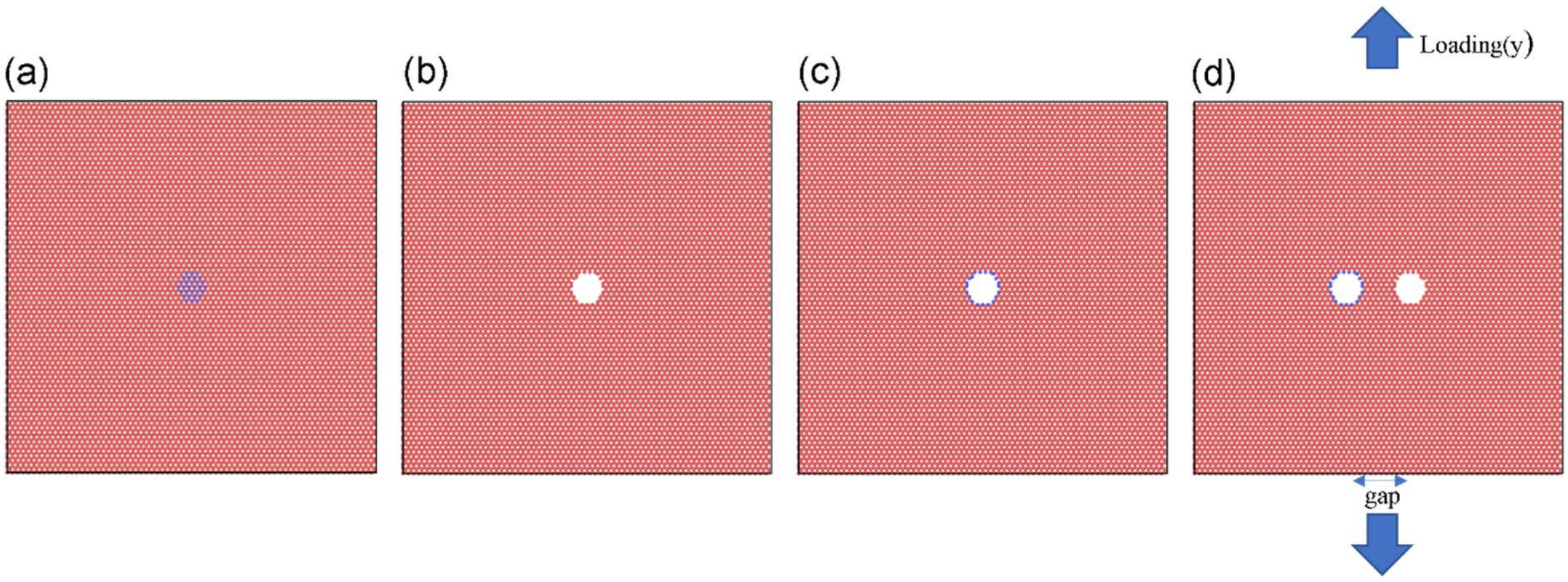

**Figure 1.** Atomic configurations of the defect models used in this study. (a) Graphitic-nitrogen cluster, (b) circular void, and (c) pyridinic-nitrogen cluster, each introduced at the centre of a graphene sheet of dimensions 169.7 × 170.4 Å (carbon shown in grey, nitrogen in blue). (d) Representative coupled configuration containing a nitrogen cluster and a void separated by an edge-to-edge gap, with the loading direction indicated.

### *2.3 Interatomic potential*

Interatomic interactions among carbon and nitrogen atoms were described using the Tersoff bond-order potential with the optimized B–C–N parameterization of Kınacı et al. [12] in which the C–C parameters derive from the lattice-dynamics-optimized set of Lindsay and Broido [13]. This bond-order form captures C–C, C–N, and N–N bonding, as well as the bond-breaking required to model fracture. The potential was used with its published cutoff parameters without modification: the C–C interaction employs a cutoff radius R = 1.95 Å and width D = 0.15 Å (switching region 1.80–2.10 Å), with R = 1.95 Å, D = 0.10 Å for C–N and R = 2.0 Å, D = 0.1 Å for N–N. We note that, like all Tersoff-family potentials, this parameterization can slightly overestimate the fracture stress and strain at large tensile strains owing to the form of the cutoff function; because the present study is comparative, properties of defective and coexisting- defect systems are referenced to pristine and single-defect cases computed identically, such systematic effects largely cancel, and the pristine values obtained here agree with established literature (See section 3.1).

### *2.4 Energy minimization, equilibration, and tensile loading*

Each configuration was first relaxed by conjugate-gradient energy minimization with energy and force tolerances of $1 \times 10^{-11}$ eV and $1 \times 10^{-6}$ eV/Å, respectively. The minimized structure was thermally equilibrated in the isothermal–isobaric (NPT) ensemble at 300 K and zero in-plane pressure for 5 ps using a Nosé–Hoover thermostat and barostat [14], [15] with a timestep of 0.5 fs. Uniaxial tensile loading was then applied along the armchair (y) direction at a constant engineering strain rate of $1 \times 10^{9}$ $s^{-1}$, with the temperature maintained at 300 K by a Nosé–Hoover thermostat and the transverse (x) direction allowed to contract freely to accommodate the Poisson effect. Reflective boundaries were imposed on the non-periodic z-surfaces to retain atoms detaching from the fracturing edges. Each configuration was simulated using three independent initial velocity distributions (random seeds) to enable statistical averaging.

### *2.5 Stress and mechanical-property extraction*

The stress tensor was computed from the virial theorem, and the engineering stress was obtained by normalizing the relevant tensor component with the instantaneous cell volume, assuming an effective graphene thickness of 3.4 Å. Stress–strain curves were generated for every configuration. The Young's modulus was obtained from the slope of the stress–strain curve in the small-strain elastic regime (0.5–2 % strain); the ultimate tensile strength was taken as the peak stress; and the fracture strain as the strain at that peak. All reported values are means over the three independent runs, with uncertainties reported as standard deviations.

### *2.6 Analysis of defect interaction and visualization*

The mechanical influence of coexisting defects was assessed by comparing the properties of the two-defect (coupled) cells with those of the isolated single-defect references—the graphitic cluster, the pyridinic cluster, and the void—and pristine graphene, all at matched defect size, as functions of the inter-defect separation and of the orientation relative to the loading axis. For the enlarged perpendicular cell, the single-defect references were regenerated in the same enlarged cell to ensure size-consistent comparison. Atomic configurations and the per-atom (von Mises) stress field during deformation and fracture were visualized and analysed using OVITO (version 3.10.6) [16].

# 3. Results

### *3.1 Validation of the simulation protocol*

The tensile response of pristine graphene was first computed to validate the simulation protocol. The pristine sheet exhibited a Young's modulus of 937 ± 5 GPa, an ultimate tensile strength (UTS) of 127.8 ± 0.7 GPa, and a fracture strain of 0.254 ± 0.002 (mean ± standard deviation over three independent runs). These values are consistent with the experimentally measured and computationally reported mechanical properties of pristine [1], [17], confirming that the present model and loading protocol reproduce the established mechanical behaviour of graphene and providing a reliable baseline for the defect analyses that follow.

**Table 1.** Mechanical properties of pristine graphene in the present work compared with experimental and computational values reported in the literature.

| Property | This work | Literature | Ref. |
|---|---|---|---|
| Young's modulus (GPa) | 937 ± 5 | ~1000 (≈1 TPa) | [1] |
| Ultimate tensile strength (GPa) | 127.8 ± 0.7 | ~130 | [1] |
| Fracture strain | 0.254 ± 0.002 | ~0.25 | [17] |

(Values are mean ± standard deviation over three independent simulations.)

### *3.2 Mechanical response of isolated defects: chemistry versus missing atoms*

To separate the mechanical roles of nitrogen chemistry and missing atoms, the three size-matched defects—a graphitic-nitrogen cluster, a circular void, and a pyridinic-nitrogen cluster—were examined in isolation and compared with pristine graphene. The resulting properties are summarized in Table 2 and Figure 2, and representative stress–strain curves are shown in Figure 3.

**Table 2.** Young's modulus, ultimate tensile strength, and fracture strain of pristine graphene and the three isolated, size-matched defects (mean ± standard deviation over three independent simulations).

| System | Young's modulus (GPa) | UTS (GPa) | Fracture strain | UTS change |
|---|---|---|---|---|
| Pristine | 937 ± 5 | 127.8 ± 0.7 | 0.254 ± 0.002 | — |
| Graphitic-N cluster | 936 ± 5 | 127.8 ± 0.7 | 0.252 ± 0.002 | — |
| Void | 916 ± 3 | 98.0 ± 2.2 | 0.158 ± 0.007 | −23 % |
| Pyridinic-N cluster | 924 ± 3 | 90.0 ± 1.8 | 0.132 ± 0.004 | −30 % |

(Values are mean ± standard deviation over three independent simulations; UTS change is relative to pristine.)

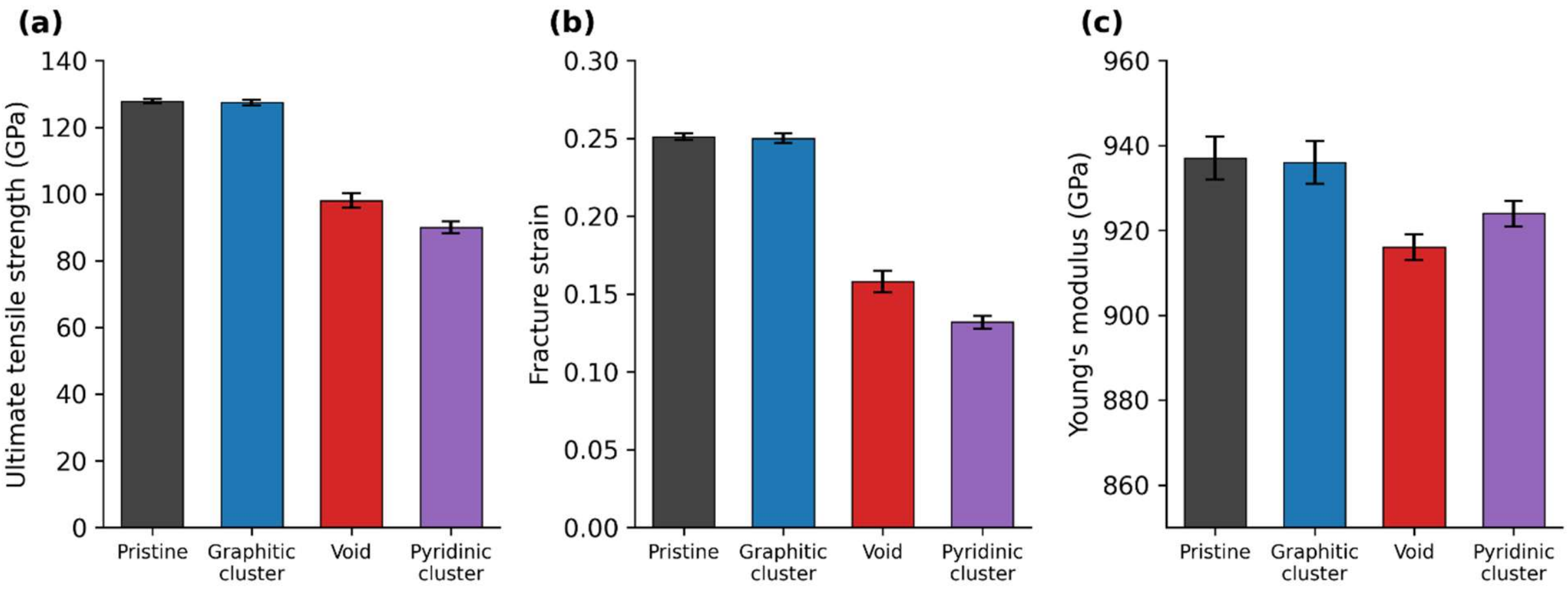

**Figure 2.** Mechanical properties of pristine graphene and the three isolated defects: (a) ultimate tensile strength, (b) fracture strain, and (c) Young's modulus. Error bars denote the standard deviation over three independent simulations.

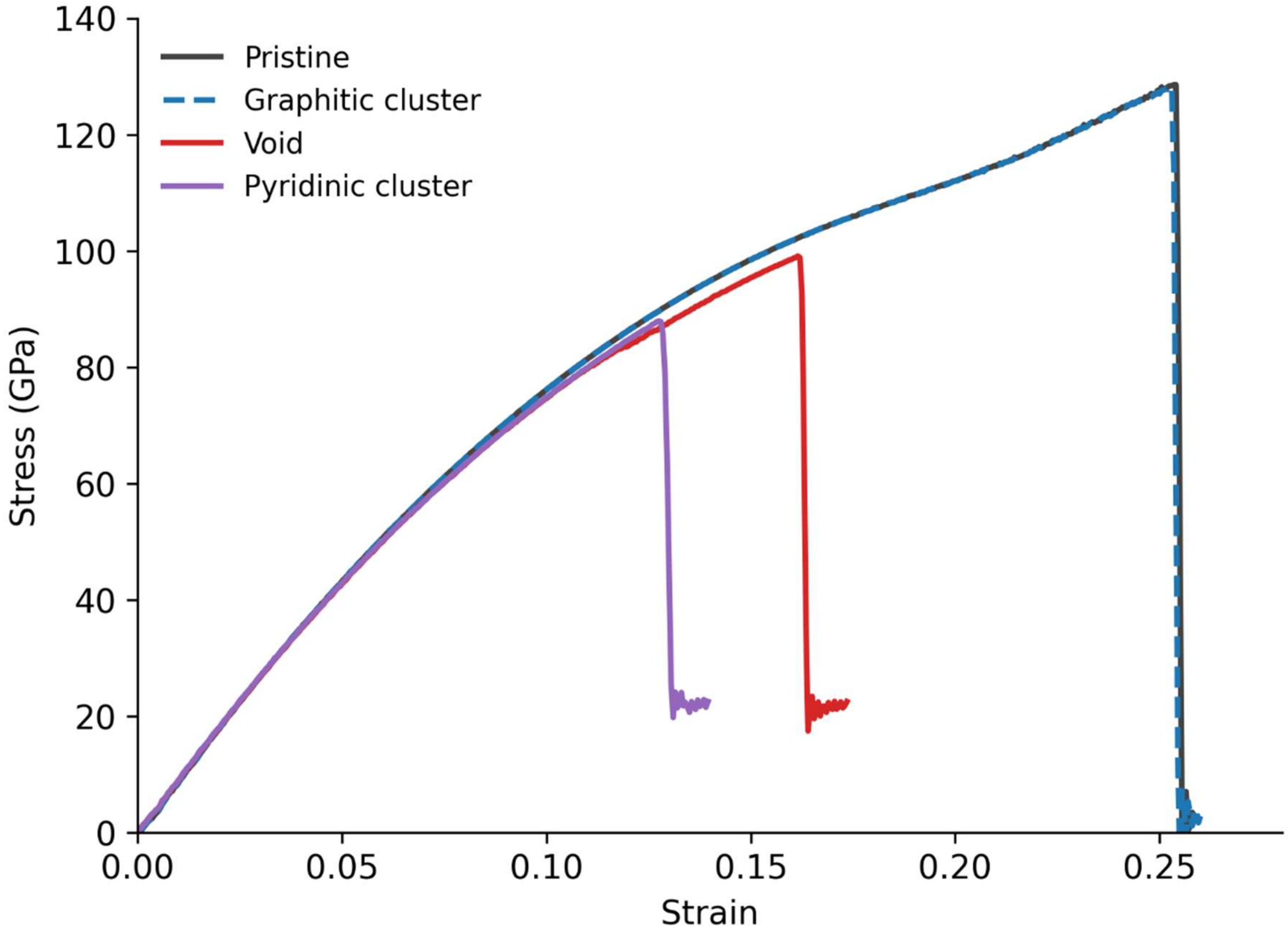

**Figure 3.** Representative stress–strain curves of pristine graphene and the graphitic-nitrogen cluster, void, and pyridinic-nitrogen cluster under uniaxial tension.

The graphitic-nitrogen cluster is mechanically almost indistinguishable from pristine graphene: its UTS (127.5 ± 0.8 GPa) and fracture strain (0.252 ± 0.003) differ from the pristine values by less than the run-to-run scatter, and its modulus is likewise unchanged. This near-invisibility follows directly from the atomic configuration: substitutional graphitic nitrogen removes no atoms, preserves the three-fold-coordinated $sp^2$ network, and forms C–N bonds of strength comparable to C–C bonds, so the lattice carries load almost as if undisturbed and develops no localized stress concentration [18]. A chemically realistic graphitic cluster therefore leaves the mechanical response of graphene essentially intact, in contrast to the pronounced degradation associated with vacancy-type defects discussed below [7], [19].

In contrast, the circular void degrades the mechanical properties markedly, reducing the UTS to 98.0 ± 2.2 GPa (a 23 % decrease) and the fracture strain to 0.158 ± 0.007 (a 37 % decrease). Removing atoms severs load-bearing bonds and creates an under-coordinated pore that concentrates stress, promoting premature bond rupture, in agreement with the established behaviour of vacancy- and hole-type defects in [20], [21], [22].

The pyridinic-nitrogen cluster is the most damaging of the three defects, with a UTS of 90.0 ± 1.8 GPa (a 30 % decrease) and a fracture strain of 0.132 ± 0.004 (a 47 % decrease). Because the pyridinic cluster and the void share an identical hole and differ only in whether the rim is terminated by carbon or by nitrogen, the comparison isolates the mechanical effect of the edge chemistry directly: the nitrogen-terminated pore is significantly weaker than the bare pore, both in strength (8.0 GPa lower, ~2.8 times the combined uncertainty) and in fracture strain (0.026 lower, ~3.3 times the combined uncertainty). The two-fold- coordinated, under-coordinated edge nitrogen softens the rim relative to a reconstructed carbon edge, lowering the strain at which the pore begins to tear [3], [23].

Taken together, these controlled comparisons yield a clear decomposition of how a nitrogen cluster affects the mechanical integrity of graphene. Nitrogen chemistry in the absence of missing atoms (the graphitic cluster) has a negligible effect; the removal of atoms (the void) is degrading; and the combination of missing atoms with edge nitrogen (the pyridinic cluster) is the most degrading. The mechanical impact of a nitrogen cluster is therefore governed almost entirely by whether it carries vacancies rather than by the presence of nitrogen itself.

### *3.3 Fracture mechanisms*

The atomic-scale failure processes, visualized through the per-atom stress field (Figure 4), clarify the origin of these differences and were found to be consistent across all three independent runs for each defect. In the void, bond rupture initiates well before the stress peak: localized damage first appears at the high-stress lobes flanking the pore at a strain of approximately 0.10, while the surrounding lattice accommodates this incipient damage so that the stress continues to rise to its maximum near a strain of 0.15, at which point the separated regions coalesce into a crack that propagates transverse to the loading direction and the stress collapses. The void thus exhibits a degree of damage tolerance, sustaining sub-critical damage over a substantial strain range before final failure.

The pyridinic cluster fails through a distinctly different, more abrupt process. The nitrogen-terminated pore remains intact up to its peak stress at a strain of approximately 0.135, with stress concentrated at the lobes flanking the pore; bond rupture then initiates at the nitrogen-decorated rim essentially at the peak and is immediately followed by brittle crack propagation and stress collapse. The pyridinic pore therefore forgoes the damage-tolerant pre-peak phase exhibited by the bare void, reaching a lower peak strain and failing abruptly once its rim gives way. This difference in mechanism accounts for the pyridinic cluster exhibiting both the lowest strength and the lowest fracture strain of the three defects: the under-coordinated edge nitrogen nucleates failure at the rim at a lower strain than the reconstructed carbon edge of the bare void.

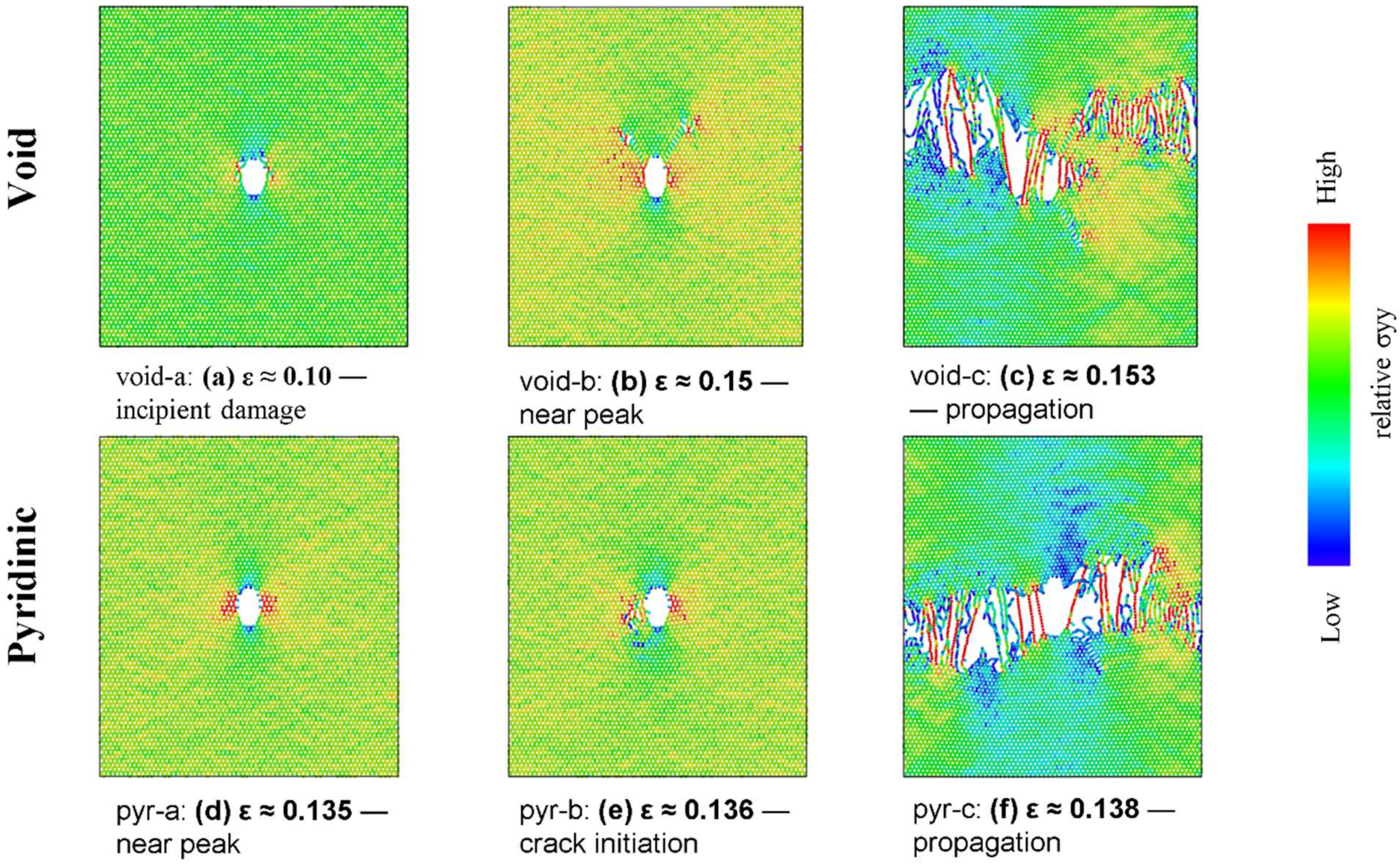

**Figure 4.** Per-atom stress fields during deformation and fracture, coloured by the stress component along the loading direction, for (top row) the void and (bottom row) the pyridinic-nitrogen cluster at successive stages of loading: (a) incipient/peak loading, (b) crack initiation, and (c) propagation.

Finally, the Young's modulus varies only weakly across all systems (916–937 GPa, a spread of ~2 %), because a single ~7 Å defect occupies a small fraction of the large simulation cell and perturbs the small-strain elastic stiffness only slightly; the failure-controlled properties (UTS and fracture strain), governed by local stress concentration, carry the mechanical signal and are emphasized accordingly.

### *3.4 Interaction between a nitrogen cluster and a void*

Real defective graphene contains multiple defects in proximity, whose perturbed stress fields may overlap and interact. To quantify this, configurations containing a pyridinic-nitrogen cluster and a void—the two mechanically active defects identified above—were examined as a function of their edge-to-edge separation and of their orientation relative to the loading axis. Two arrangements were considered: parallel, with the defects offset along the loading direction (in-line), and perpendicular, with the defects offset across the loading direction (side-by-side). The resulting strengths are shown in Figure 5.

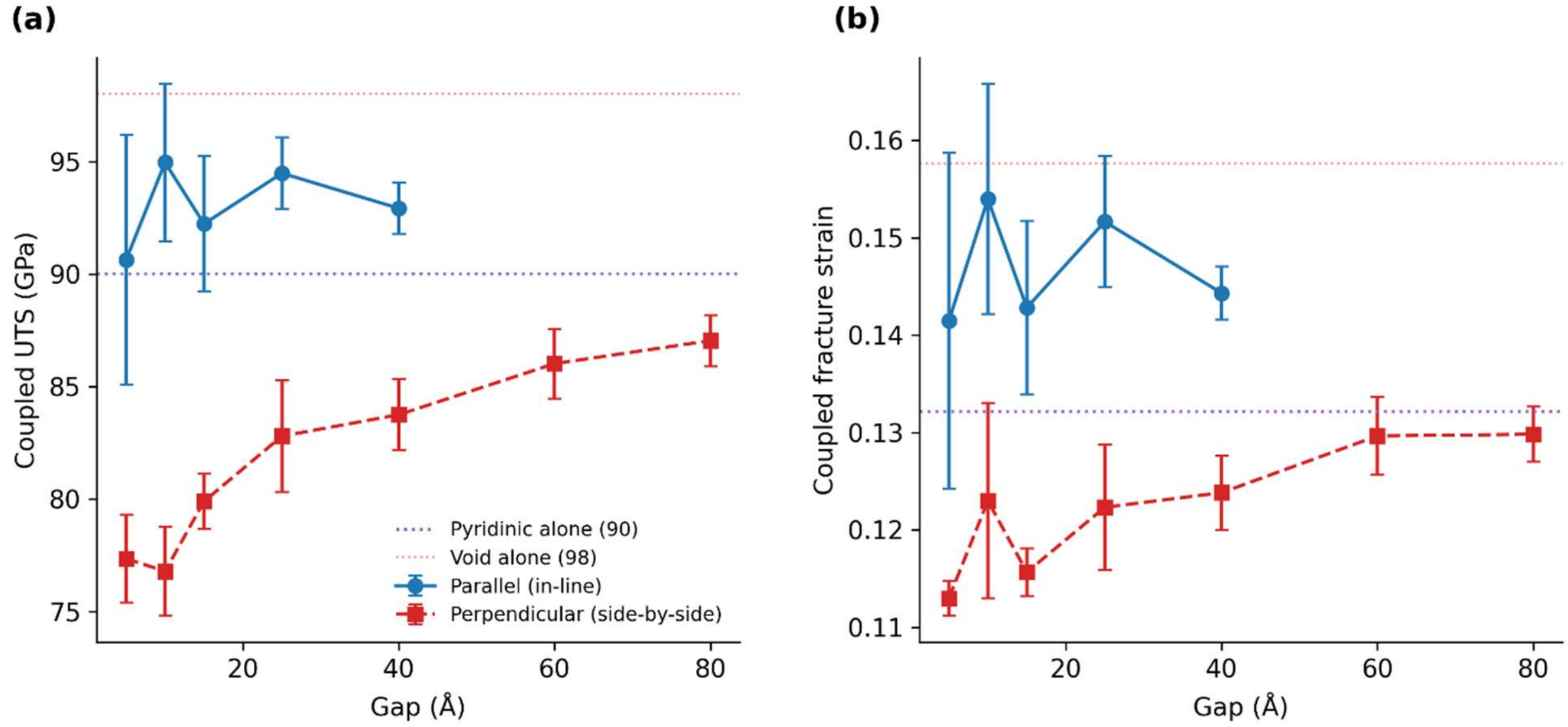

**Figure 5.** Coupled (a) ultimate tensile strength and (b) fracture strain of a nitrogen cluster and a void as a function of edge-to-edge gap, for the parallel (in-line) and perpendicular (side-by-side) orientations. Dotted lines indicate the corresponding single-defect reference values.

The mechanical interaction is governed overwhelmingly by orientation. For every separation studied, the perpendicular (side-by-side) arrangement is substantially weaker than the parallel (in-line) arrangement, by approximately 13 GPa in UTS. This difference reflects the load-bearing geometry: when the two defects lie on the same cross-section normal to the loading axis (perpendicular), that cross-section is reduced by both defects simultaneously and fails early, whereas when they are stacked in-line (parallel) each loaded cross-section contains at most one defect, so the minimum load-bearing section is less severely reduced.

In the parallel orientation the coupled strength is essentially independent of separation and coincides with the strength of the pyridinic cluster alone (~90 GPa) across all gaps. The in-line pair therefore behaves as a weakest-link system: failure is dictated by the more severe of the two defects—the pyridinic cluster—and the void, lying in a different cross-section, contributes little. No significant proximity interaction is observed in this orientation.

The perpendicular orientation, by contrast, exhibits a genuine, separation- dependent interaction. The coupled strength falls below that of either isolated defect and decreases progressively as the defects approach, from 87.0 ± 1.1 GPa at a gap of 80 Å to 77.4 GPa at 5 Å; the coupled fracture strain follows the same trend. As the two side-by-side stress concentrators are brought together, their perturbed stress fields increasingly overlap, intensifying the stress in the ligament between them and lowering the strength of the shared cross-section.

To establish the range of this interaction, the perpendicular separation was extended to 60 and 80 Å, with the single-defect references regenerated in the correspondingly enlarged cell to ensure a size-consistent comparison; the enlarged-cell references reproduced the standard-cell values to within the statistical uncertainty. At a gap of 80 Å the coupled strength (87.0 ± 1.1 GPa) recovers to within ~2 GPa of the pyridinic cluster alone (89.2 GPa), indicating that the interaction has effectively decayed to the non-interacting (weakest-link) limit. The side-by-side interaction is thus comparatively long-ranged, remaining measurable at separations of several nanometres and converging to the single-defect limit only by ~80 Å.

These results indicate that, for coexisting defects in graphene, the relative geometry is more important than proximity per se: defects aligned with the load interact negligibly and the assembly fails at its weakest member, whereas defects arranged across the load couple through their overlapping stress fields and progressively weaken the sheet as they approach.

## 4. Discussion

The results above converge on a single organizing principle: the mechanical integrity of nitrogen-modified graphene is governed by the configuration of its defects—whether nitrogen is accompanied by missing atoms, and how coexisting defects are arranged relative to the loading direction—rather than by the amount of nitrogen present. This reframes the mechanical role of nitrogen in graphene from a question of composition to one of local atomic structure.

The single-defect comparisons make this explicit. A graphitic-nitrogen cluster, in which nitrogen substitutes for carbon without disrupting the three-fold- coordinated network, leaves strength, stiffness and ductility essentially unchanged, whereas a vacancy of the same footprint degrades all of them substantially. Decorating that same vacancy with edge nitrogen—the pyridinic configuration—degrades the properties further still. Because real nitrogen-doped graphene contains graphitic, pyridinic and pyrrolic nitrogen in proportions that depend strongly on synthesis conditions [24], this distinction has direct practical consequences: the mechanical penalty of nitrogen doping is carried almost entirely by the vacancy-associated (pyridinic, pyrrolic) fraction, while the graphitic fraction is mechanically benign. Two samples with identical total nitrogen content could therefore differ markedly in strength depending on how that nitrogen is bonded. This may help reconcile the scatter in reported mechanical properties of nitrogen-doped graphene, and it suggests that mechanically non-destructive functionalization is best achieved through graphitic substitution.

The fracture mechanisms underlying these differences are themselves instructive. The bare void tolerates sub-critical damage: bond rupture begins at its stress-concentrating flanks well below the peak stress, yet the surrounding lattice redistributes load so that the sheet continues to strengthen until the damage coalesces. The pyridinic pore, by contrast, remains intact to a lower peak and then fails abruptly once its nitrogen-terminated rim gives way. The under-coordinated edge nitrogen thus acts as a premature nucleation site, exchanging the modest damage tolerance of the carbon edge for earlier, more brittle failure. This mechanistic contrast—not merely the difference in peak values—is what distinguishes a chemically decorated pore from a bare one, and it underscores that edge chemistry, not nitrogen content per se, controls how a nitrogen cluster fails.

The behaviour of coexisting defects adds a second, orthogonal design variable: geometry. The interaction between a nitrogen cluster and a void is dominated not by their separation but by their orientation relative to the load. Defects arranged in-line with the loading axis interact negligibly, and the pair fails at its more severe member in a weakest-link fashion; defects arranged across the load share a common load-bearing cross-section and couple through their overlapping stress fields, weakening the sheet progressively as they approach. This side-by-side interaction is comparatively long-ranged, persisting to separations of several nanometres and decaying to the single-defect limit only near 80 Å. The implication for real, multiply-defective graphene is that strength is set less by defect density alone than by the spatial arrangement of defects relative to the applied stress: a population of defects aligned across the loading direction is considerably more detrimental than the same defects distributed in-line, even at identical density. Conventional descriptions based on average defect concentration therefore cannot, by themselves, capture the mechanical response of defective graphene.

Taken together, these findings outline a route toward defect-tolerant design. Where nitrogen functionalization is desired for electronic, catalytic or chemical purposes, graphitic substitution offers those modifications at negligible mechanical cost, whereas vacancy-associated nitrogen and the clustering or cross-load alignment of defects should be minimized where mechanical reliability is paramount. The mechanical consequences of doping and of defect coexistence are, in this view, controllable through configuration rather than being fixed penalties of disorder.

Several limitations should be borne in mind. The simulations employ a single empirical bond-order potential; while it captures bond breaking and was used here in a strictly comparative manner with all systems referenced to identical pristine and single-defect baselines, Tersoff-type potentials are known to represent the large-strain fracture regime only approximately, and absolute strengths and fracture strains should be interpreted with that caveat [25]. The defects considered are idealized—circular voids and compact clusters of fixed size and uniform edge termination—whereas real defects exhibit a range of shapes, sizes, reconstructions and mixed edge chemistries. The imposed strain rate (10^9 s^-1) and temperature (300 K) are typical of molecular dynamics but far removed from most experimental conditions, and strain rate is known to influence the computed strength and ductility [26]. Finally, the sheets are freestanding and single-layer; substrate interactions, multilayer stacking and out-of-plane rippling are not considered. These factors bound the quantitative claims but do not affect the comparative, configuration-based conclusions that are the focus of this work.

These limitations also indicate productive directions for future work. Machine- learning interatomic potentials trained on first-principles fracture data could place the absolute values on a firmer footing while retaining access to the large system sizes required to study defect interactions [27]. Extending the present configuration-based analysis to realistic, statistically distributed defect populations would connect atomistic mechanisms to the continuum strength of defective graphene. Perhaps most importantly, because the same structural defects that govern mechanical behaviour also strongly modulate thermal and electronic transport, a unified treatment of how defect configuration simultaneously controls mechanical, thermal and electronic response —particularly in two-dimensional heterostructures[28], where interfaces and moiré superlattices introduce additional tunable degrees of freedom [29]—represents a natural and important extension of this study. More broadly, the present configuration-based framework, in which mechanical degradation is decomposed into chemistry, vacancy, and geometric-arrangement contributions, may extend naturally to radiation-induced defect populations relevant to aerospace and space-material applications, where displacement damage similarly couples missing-atom and geometric effects.

## 5. Conclusions

Using molecular dynamics simulations of uniaxial tension, this work has examined how nitrogen-containing defects and their coexistence govern the mechanical behaviour of graphene, with the aim of separating the roles of nitrogen chemistry, missing atoms, and defect arrangement. Three principal conclusions emerge.

First, the mechanical effect of a nitrogen cluster is controlled by whether it carries vacancies, not by the nitrogen itself. A graphitic-nitrogen cluster, which preserves the intact $sp^2$ network, leaves the strength, stiffness and fracture strain of graphene essentially unchanged, whereas a void of equal size degrades them substantially, and a pyridinic cluster—the same void decorated with edge nitrogen—is the most damaging of the three. Because the pyridinic cluster and the void differ only in edge chemistry, and the graphitic and pyridinic clusters differ only in the presence of the vacancy, these controlled comparisons isolate the two contributions and show that substitutional nitrogen is mechanically benign while vacancy-associated nitrogen is not.

Second, these defects fail by distinct mechanisms. The bare void sustains sub-critical damage at its stress-concentrating flanks well below its peak stress and retains a degree of damage tolerance before final, transverse fracture, whereas the nitrogen-decorated pyridinic pore remains intact to a lower peak and then fails abruptly from its rim. The under-coordinated edge nitrogen thus nucleates failure earlier than a reconstructed carbon edge, accounting for the pyridinic cluster exhibiting both the lowest strength and the lowest fracture strain.

Third, when a nitrogen cluster and a void coexist, their mechanical interaction is governed primarily by orientation relative to the load. Defects aligned in-line with the loading axis interact negligibly and the pair fails at its more severe member, whereas defects arranged across the load couple through their overlapping stress fields and weaken the sheet progressively as they approach; this side-by-side interaction is comparatively long-ranged, decaying to the single-defect limit only near a separation of 80 Å.

Taken together, these results establish that the mechanical integrity of nitrogen-modified graphene is determined by defect configuration—the bonding environment of the nitrogen and the geometric arrangement of coexisting defects relative to the load—rather than by nitrogen content or defect density alone. This configuration-based view offers a basis for defect-tolerant design, in which graphitic substitution provides chemical functionalization at negligible mechanical cost while vacancy-associated nitrogen and cross-load defect alignment are avoided where reliability is critical. Extending this framework with machine-learning interatomic potentials and to the coupled mechanical, thermal and electronic response of defective two-dimensional materials represents a promising direction for future work.

**Funding**

No funding was used to do this research work.

**Data availability**

Data will be made available on request.

**Declarations**

**Conflict of interests:** The authors declare no financial or non-financial competing interests that could have appeared to influence the work reported in this paper.

**Credit authorship contribution statement:**

**Indranil Rudra:** Conceptualization, Methodology, Software, Investigation, Formal analysis, Visualization, Writing – original draft. **Jahid Emon:** Supervision, Writing – review and editing. **A.K.M. Monjur Morshed:** Supervision.